\documentclass[letter]{aa}       
\usepackage{graphicx}
\usepackage{amssymb}
\usepackage{amsmath}
\usepackage{longtable}
\usepackage{supertabular}
\usepackage{natbib}
\bibpunct{(}{)}{;}{a}{}{,}

\defcitealias{Kad08}{K08}

\begin{document}

\title{On the nature of an ejection event in the jet of 3C~111}

\author{M. Perucho\inst{1} \and
        I. Agudo \inst{2} \and
        J.~L. G\'omez \inst{2} \and
        M. Kadler\inst{3,4,5} \and
        E. Ros\inst{1} \and
        Y. Y. Kovalev\inst{1,6}
    }

\authorrunning{Perucho et al.}

\titlerunning{On the nature of an ejection event in the jet of 3C~111}

\institute{Max-Planck-Institut f\"ur Radioastronomie, Auf dem H\"ugel 69, 53121 Bonn, Germany 
\and
Instituto de Astrof\'{\i}sica de Andaluc\'{\i}a (CSIC),  Apartado 3004, E-18080 Granada, Spain 
\and 
Dr. Karl Remeis-Observatory, University of Erlangen-Nuremberg, Sternwartstrasse 7, 96049 Bamberg, Germany \and 
CRESST/NASA Goddard Space Flight Center, 662 Greenbelt, MD 20771, USA 
\and 
Universities Space Research Association, 10211 Wincopin Circle, Suite 500 Columbia, MD 21044, USA 
\and
Astro Space Center of Lebedev Physical Institute, Profsoyuznaya 84/32, 117997 Moscow, Russia}

\offprints{M. Perucho, \email{perucho@mpifr-bonn.mpg.de}}

\date{Received <date> / Accepted <date>}

\abstract{
{We present a possible scenario for the ejection of a superluminal component in the jet of the Broad Line Radio Galaxy \object{3C~111} in early 1996. VLBI observations at 15 GHz discovered the presence of two jet features on scales smaller than one parsec. The first component evolves downstream, whereas the second one fades out after 1 parsec.}
{We propose the injection of a perturbation of dense material followed by a decrease in the injection rate of material in the jet as a plausible explanation.}
{This scenario is supported by 1D relativistic hydrodynamic and emission simulations. The perturbation is modeled as an increase in the jet density, without modifying the original Lorentz factor in the initial conditions.}
{We show that an increase of the Lorentz factor in the material of the perturbation fails to reproduce the observed evolution of this flare. We are able to estimate the lifetime of the ejection event in 3C~111 to be $36\pm7$~days.} {}

}

\maketitle

\section{Introduction} \label{intro}
Flaring events at radio frequencies are known to take place in
Active Galactic Nuclei (AGN), usually followed by the observation
of new radio features in the parsec-scale jets \citep[e.g.,][]{sa02}. It has been shown that the ejection of those features, or components, is related to dips in the X-ray emission from the active nucleus in the case of 3C~120 \citep{Mar02}, and perhaps also in 3C~111 \citep{Mar06}. The dips in X-rays precede the observations of new radio-components. The decrease in X-ray emission may be caused by the loss of the inner regions of the disc. In this scenario, a fraction of the accreted material is injected in the jet and a new component is later observed in VLBI images, after the material becomes detectable at the observing frequencies, as it evolves downstream. The components are interpreted as the shocks produced by the ejection of denser and/or faster plasma in the flaring event from the accretion disc \citep{mar85}. The conditions for triggering the ejection of the material in those radio features are still unknown. In the case of microquasars, it has been proposed that the stronger components are ejected right before the passage of the source from the X-ray hard/low state, associated with higher radio brightness, to the soft/high state, associated with lower radio emission -- a decrease in the injection of emitting particles in the jet \citep[][ and references therein]{fb04}.

If we interpret jet components as shocks propagating in a supersonic flow, their origin must be related to an increase of pressure and/or velocity in an injected perturbation with respect to the steady initial flow. Hydrodynamical simulations \citep[][ A03 hereafter]{Alo03} have shown that such 
jet perturbations produce a forward and a reverse structure, which would be expected to be observed as a fast front and a slower back component. 

\begin{figure}[!t]
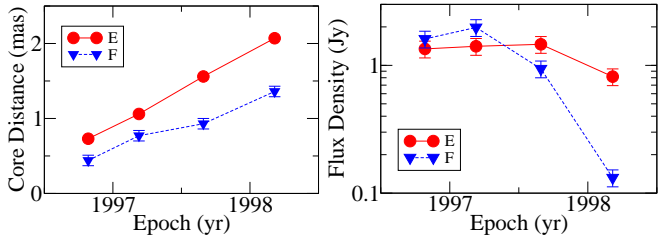

   \centering
\includegraphics[clip,angle=0,width=0.17\textheight]{fig1a}
\includegraphics[clip,angle=0,width=0.18\textheight]{fig1b}
\caption{Core distance and flux density evolution with time of components E and F in 3C~111, based on the results from \citetalias{Kad08}.}
\label{fig:0}
\end{figure}

In the jet of the Broad Line Radio Galaxy (BLRG) 3C~111 ($z=0.049$, $1\,\rm{mas}\simeq 1\,\rm{pc}$), a very strong flaring event in early 1996 gave rise to the ejection of two jet
features observed at 15 GHz with the Very Long Baseline
Array (labeled as components E and F -- see Fig.~\ref{fig:0} and \citealt*{Kad08}, hereafter \citetalias{Kad08}). Both component trajectories can be back-extrapolated to similar ejection epochs within 3 months (around 1996.10). However, they show different speeds and the time evolution of their brightness is different (see Fig.~\ref{fig:0}): the inner component F is initially brighter (1996.82 and 1997.19) and fades out very rapidly (1997.66 and 1998.18), while the leading component E shows a slower decrease in flux density. After 1999 (see \citetalias{Kad08}),
F disappeared and E evolves, accelerating
and generating trailing components in its wake \citep{Agu01}. The differences betweeen E and F cannot be attributed to different Doppler factors of the components, as these are very similar ($D_E\sim3.2$, $D_F\sim3.1$, following \citetalias{Kad08}). The possibility that component F represents a second injection after component E is highly improbable on the basis of its velocity and its brightness evolution, which cannot be linked to the propagation in the wake of the latter.
In this letter, we investigate, in a qualitative way, the possibility that these components are the front and rear region of a single perturbation. A quantitative comparison between the simulations and the observations would require detailed knowledge about the nature of the flow and is out of the scope of this work. The numbers and error estimates used in this letter that are related to the observations are taken or derived from \citetalias{Kad08}.

\section{Hydrodynamics and emission}\label{sec:pert}
In numerical simulations, the enhanced injection of material in
jets has been modelled as an abrupt square perturbation of the
flow density and/or Lorentz factor at the injection point \citep[e.g., ][]{go97,Agu01} in a
steady jet that recovers the initial state after the perturbation is completely ejected. 
A03 show that the perturbation generated by the injection of denser plasma propagates
downstream, spreading in the axial direction along the jet
and finally splitting into two distinct regions.
The forward region moves against the underlying flow and the
reverse region propagates backwards in the jet fluid reference
frame, thus acting against the fluid that is injected after the perturbation. The 
injection of dense fluid behind the perturbation feeds the reverse shock, making it potentially observable (with similar brightness to that of the forward shock, A03) for relatively long times. 

\begin{figure*}[!t]
   \centering
\includegraphics[clip,angle=0,width=\linewidth]{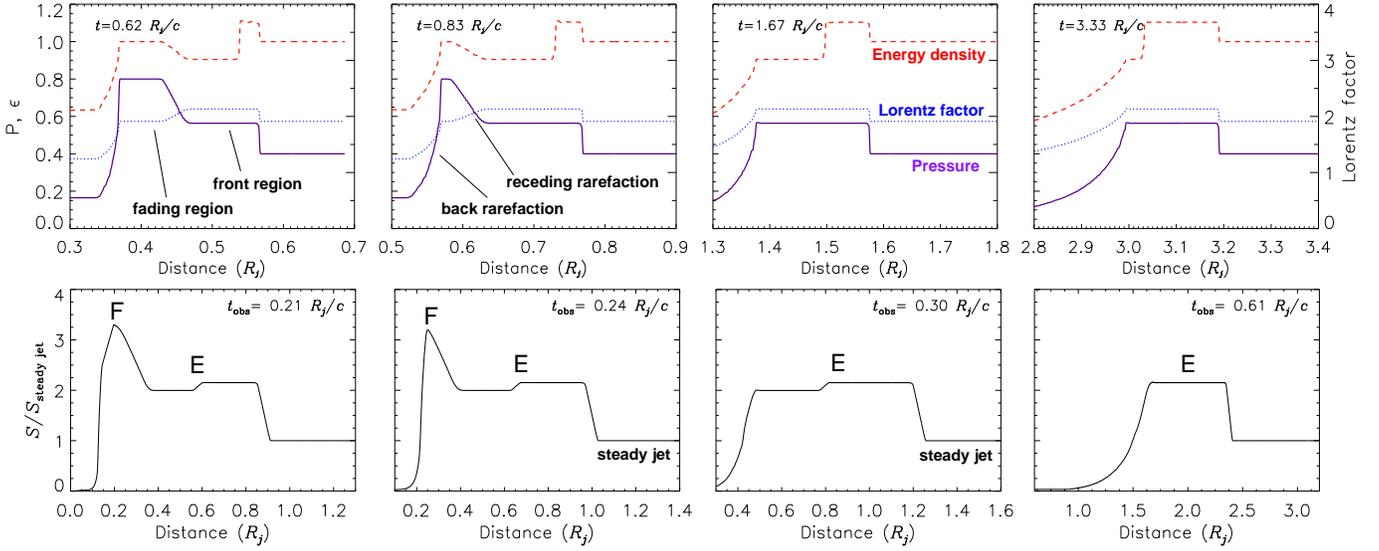}
\caption{
Snapshots of the evolution (left to right) of a square perturbation
injected in a steady jet, followed by a strong rarefaction. The
dotted-light-blue lines stand for the Lorentz factor, the solid-dark-blue
line for pressure and the dashed-red lines for specific internal energy. 
The simulation is run with 24000~cells; the velocity of the initial flow is $v_j=0.9\,c$; the perturbation is injected during a time interval $0.2\,R_j/c$, with a density twice that of the jet, the same specific internal energy and velocity $v_p=0.9\,c$; the rarefied medium is injected after the perturbation with the same velocity as the initial flow, and a pressure ten times
smaller than that of the initial flow. Please note the change of scale in the abcissae. The bottom panels show the simulated total intensity emission along the jet axis at four representative epochs. The identification of the features in the simulation with the observed
components E and F in \citetalias{Kad08} is indicated in each panel. A jet width of 500 cells and axial symmetry is used to compute the emission.}
\label{fig:1}
\end{figure*}

We have performed one-dimensional numerical relativistic hydrodynamic (RHD) simulations in
which a square perturbation in density is injected into a steady jet, without modifying the initial Lorentz factor, and relaxing the condition that the initial jet flow is reestablished immediately after the perturbation. We have substituted this by a rarefied flow, representing a reduction of the injection rate, in order to avoid the formation of a strong reverse shock. In this picture, the original jet injection rates should be recovered after some time. However, in this work we only focus on the evolution of the strong ejection and the period before the reestablishment of the jet flow. Multidimensional simulations are out of the scope of this work due to the computational effort required and to the one-dimensional character of this problem. The simulations have been performed using a numerical code that solves the equations of relativistic hydrodynamics written in the conservation form, as described in \cite{pe05} and \cite{mart97}. The details of the simulation are given in the caption of Fig.~\ref{fig:1}. The top panels in Fig.~\ref{fig:1} show different snapshots of the evolution of the
square perturbation injected in a steady flow, in pressure, Lorentz factor and specific internal energy. Using the RHD simulations as input, we have computed the
corresponding 1D optically thin radio synchrotron emission as seen by an
observer with a line of sight at $19^{\circ}$ to the jet axis \citepalias{Kad08}. For these
computations, we used the numerical code and the procedure described in
\cite{go97} and references therein. This code takes into
account all the relevant relativistic effects, including the light
travel time delays.

In the simulation (see Fig.~\ref{fig:1}), the \textit{front} region includes the leading part of the perturbation and is identified with component E in \citetalias{Kad08}, whereas we define the \textit{fading} region as the rear part of the perturbation and identify it with component F (see Fig.~\ref{fig:1}). The material in the front region, consisting of shocked material from the steady jet and rarefied material from the perturbation separated by a contact discontinuity, shows smaller values for the pressure, and some acceleration due the propagation in the lower pressure steady jet fluid. The material in the fading region crosses the receding rarefaction that separates it from the front region (top panels in Fig.~\ref{fig:1}) and it is also ``eroded'' by the back rarefaction. Consequently, the front structure evolves, increasing its size as the front shock incorporates material from the steady jet and the material from the fading region crosses the receding rarefaction. Thus, the front region consists of the forward shock structure of the perturbation (E in Fig.~\ref{fig:1}), and the fading region is formed by the remains of the perturbation that have not crossed the receding rarefaction (F). The synchrotron emissivity (bottom panel in Fig.~\ref{fig:1}) is governed by the jet pressure and hence the emission evolution is very similar to the pressure evolution of the RHD simulations. In the emission results, the front region (component E) propagates without much flux density evolution after injection. However, the fading structure (component F), which initially shows a notably larger flux density than component E, rapidly decreases in emission as the receding and back rarefactions erode it. The reverse shock (see A03) is neither relevant nor observationally significant in our simulations, as it propagates in a very rarefied medium. For this reason it is not shown in Fig.~\ref{fig:1}.

Notice that the Lorentz factor values in Fig.~\ref{fig:1} are those corresponding to the fluid. In contrast, VLBI observations provide us with pattern velocities. In the simulation, the velocity of the front shock is measured to be $v_{s}\sim 0.96\,c$ ($v_{\rm{E}}^{obs} \sim 3.5 c$), whereas that of the fading region is $v_{r}\sim 0.87\,c$ ($v_{\rm{F}}^{obs} \sim 1.7 c$), both similar to those found in the observations \citepalias{Kad08}. The velocity of the material in the fading region is faster than that of the receding rarefaction (cf. Fig.\ref{fig:1}), as expected from the explanation in the previous paragraph. We also point out that the dilute material shown in Fig.~\ref{fig:1} presents a modified velocity due to passage through the reverse shock.

\begin{figure}[!t]
     \centering
     \includegraphics[clip,angle=0,width=\linewidth]{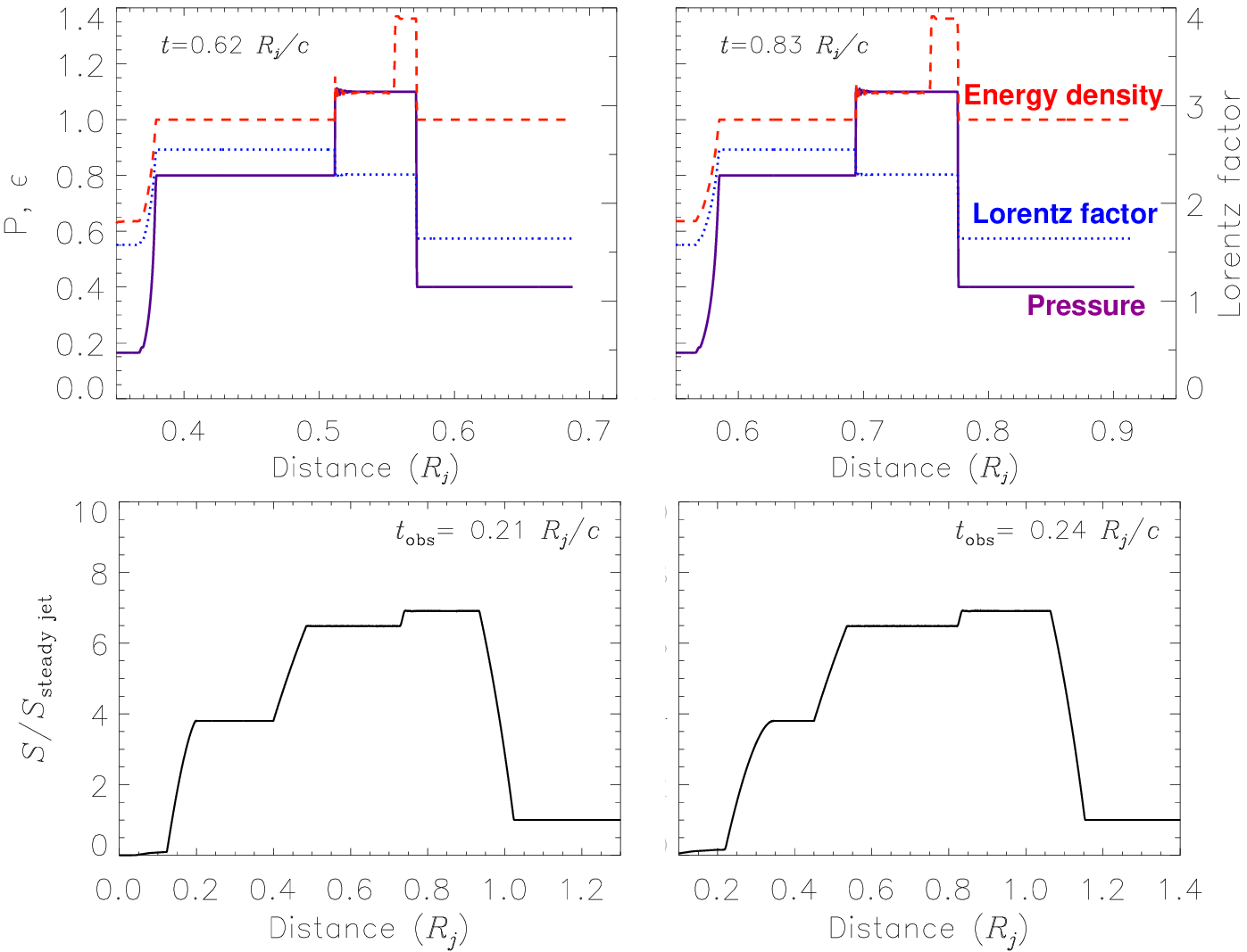}\\ 
  \caption{
Same as Fig.~\ref{fig:1} but showing the evolution of a fast square perturbation injected in a steady jet followed by a strong rarefaction (top panels) and representative emission plots (bottom panels). The perturbation is injected with a density twice that of the jet and Lorentz factor $\Gamma=3.6$. This image compares with the first two snapshots in Fig.~\ref{fig:1} and shows that, in this case, component F would not be brighter than component E. The later stages of evolution in this simulation are very similar to those in Fig.~\ref{fig:1} and are not shown here.}
  \label{fig:2}
  \end{figure}

A second simulation was performed for a faster perturbation, with Lorentz factor $\Gamma=3.6$, while keeping the rest of the parameters as in the previous simulation. The results (see Fig.~\ref{fig:2}) show that the front region of the perturbation is overpressured with respect to the rear region and, thus, the former is brighter than the latter, as shown by the emission simulations (bottom panels in Fig.~\ref{fig:2}). This is in clear contradiction to the observations of the jet in 3C~111 (Fig.~\ref{fig:0} and \citetalias{Kad08}). The difference is due to the presence of a stronger front shock. It is also important that the wave separating both regions is now a reverse shock, instead of the receding rarefaction shown in Fig.~\ref{fig:1}. This is a general result for fast perturbations, including the case of a fast perturbation in pressure equilibrium with the steady jet.

In any of the scenarios given above, the perturbed regions have enhanced emission with respect to the underlying jet. However, only in the case of an overpressured perturbation with the same Lorentz factor as the underlying flow, and including the presence of a rarefaction behind the perturbation, does the second component fade out rapidly and then the first one dominates the emission, as observed for components E and F in 3C111.

\section{Discussion and conclusions}
We are able to describe the evolution of the leading components of a major ejection in the BLRG 3C~111 as a perturbation of dense (overpressured) material followed by a dilute medium. The qualitative explanation given in the previous section supports the picture of the ejection of perturbations of dense material giving rise to radio components. Our model would remain valid if the amplitude or time-duration of the perturbation is changed. In addition, we postulate that these ejections could be followed by a decrease in the injection rate of bulk flow particles. This avoids the formation of a reverse shock, which would lead to a qualitatively different observational result. Observational support for the inclusion of this tenuous material in the simulations can be found in the prominent emission gap following behind the E/F complex in 3C~111 (see \citetalias{Kad08}). New ejection of emitting material is detected on the time scale of more than 2 years, corresponding to a gap width of up to 2\,mas in 1999 (cf. \citetalias{Kad08}). The observations of microquasar jets show that, in general, major ejections are followed by a decrease in radio brightness \citep{fb04}. If we interpret this as a decrease in the injection rate in the jet, while the system relaxes back to the initial steady state or generates the conditions for the injection of a new perturbation, it could be a process similar to that explained here. However, the strong decrease in radio emission observed in microquasar jets is not observed in AGN jets. This setup, in which we put in relation the processes taking place in the jet and the accretion disk, relies also on the results of multiwavelength observational campaigns by \cite{Mar02} and \cite{Mar06}. The latter work showed a relationship between dips in the X-ray emission from the accretion disk and the ejection of radio components in 3C~120, and possibly in 3C~111.

We can place an upper limit for the lifetime of the ejection event in its passage through the
radio core (opaque and compact emitting region at the origin of the radio jet) using only observational data. We consider that: a) the perturbation started to cross the radio core in the observed jet in 1996.10 \citepalias{Kad08}, b) the receding rarefaction ``eroding'' component F moves with the observed 
velocity $v_F=0.91\,c$ (Fig.\ref{fig:0}), and c) the last epoch at which component F is observed 
(1998.18) is taken as the time at which the receding rarefaction has completely eroded 
this component -- which is justified since the flux of component F is one order of magnitude smaller than that of component E in this epoch \citepalias{Kad08}. With these assumptions we can calculate the time that the last material of the perturbation needs to catch up the receding rarefaction, and from that, we can estimate the time lapse of the crossing through the core. We use the velocity of component E as an upper limit for the velocity of the fluid. The result tells us that the crossing of gas through the core had to end as soon as $(1.98\pm0.02)$~years before epoch 1998.18, this is, $(0.10\pm0.02)$~years after 1996.10. 
This means $\Delta t \sim (36\pm7)$~days between the passage of the
first material and that of the last portion of gas in the
perturbation. This lapse of time should be smaller if the gas
catching up the rarefaction is slower than the velocity of the head. This represents a first order estimate, which only depends on the ratio between the velocity of the material in the perturbation and that of the rarefaction wave, and provides an upper limit for the duration of such an event. We could consider this lifetime of the ejection as an upper limit for its triggering event in the black-hole/inner-accretion-disc system, as we are not taking into account the collimation and acceleration processes, which should be kinematically important in the most compact scales. In the case of microquasar jets, the radio flares that follow the dips in X-rays --and are related to the ejection of components-- have been
estimated to last between seconds and tenths of minutes \citep{mr99},
i.e., $10^4-10^6$ times smaller time-scales than for 3C~111, whereas the ratio of
the black-hole masses is of the order of $10^8$ \citep{mr99,gr06}. If we
take into account that our result represents an upper limit for the
timescale of the ejection event, this suggests that the time-scale factor between quasars and microquasars may not come directly from the ratio of black hole masses \citep[see also ][]{Mar06b}.

Strong radio flares associated with the ejection of radio-components should be carefully followed up, with a sufficiently dense time sampling, in 3C~111 and other sources for which we can achieve similar or better linear resolution with the VLBI technique. This would help in performing analyses like that presented here and to test our conclusions. The stretching of the size of any relativistic structure propagating through a jet must largely favour the detection of such double structures in the jets of nearby AGN. Databases provided by monitoring programmes such as the MOJAVE/2~cm~VLBA \citep{K04,LH05} survey are very valuable in this context. Future work includes monitoring in radio/X-ray campaigns of different AGN sources and further numerical calculations including multi-dimensional RHD, RMHD and emission simulations.

\begin{acknowledgements}
MP acknowledges support from a postdoctoral fellowship of the
``Generalitat Valenciana'' (``Beca Postdoctoral
d'Excel$\cdot$l\`encia''). IA is supported by an I3P contract with the Spanish ``Consejo Superior de Investigaciones Cient\'{i}ficas''. MP acknowledges support by
the Spanish ``Ministerio de Educaci\'on y Ciencia'' and the European Fund
for Regional Development through grants AYA2007-67627-C03-01 and AYA2007-67752-C03-02 and IA and JLG through grant AYA2007-67627-C03-03. MK has been supported by the NASA Postdoctoral Program at the Goddard Space Flight Center, administered by the Oak Ridge Associated Universities through a contract with NASA. YK is a research fellow of the Alexander von Humboldt Foundation. The authors thank J.M. Mart\'{i}, A.P. Marscher and M.A. Aloy for useful discussion and comments. The authors thank C. Fromm for his help with figures.
\end{acknowledgements}


\end{document}